\documentclass[journal=nalefd,manuscript=letter]{achemso}
\usepackage[T1]{fontenc}
\usepackage{amsmath,amssymb}
\usepackage{bm}

\newcommand{\cvs}{\text{cm}^2/ \text{V}\text{s}}

\graphicspath{ {./}{./figs/} }

\title{Fractional-Quantum Ferroelectrics: A Route to High-Mobility Ferroelectric Semiconductors}

\author{Rong-Tian Pang}
\author{Wenjie Hu}
\author{Youning Liu}
\author{Jin-Jian Zhou}
\email{jjzhou@bit.edu.cn}
\affiliation[Beijing Institute of Technology]
{Centre for Quantum Physics, Key Laboratory of Advanced Optoelectronic Quantum Architecture and Measurement (MOE), School of Physics, Beijing Institute of Technology, Beijing 100081, China.}
\alsoaffiliation[International Center for Quantum Materials]
{International Center for Quantum Materials, Beijing Institute of Technology, Zhuhai 519000, China.}


\begin{document}

\begin{tocentry}
  \centering
  \includegraphics[width=\linewidth]{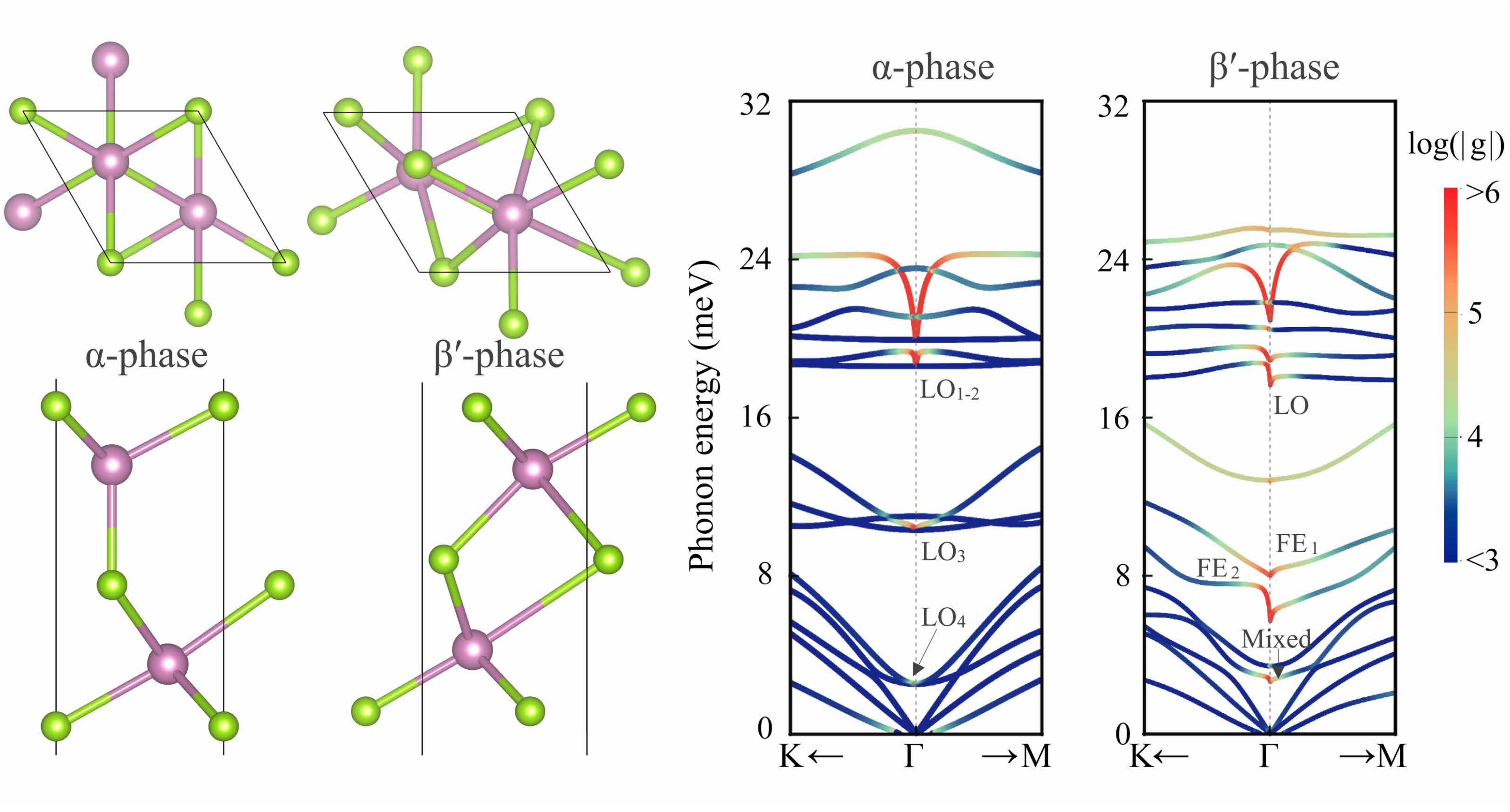}
\end{tocentry}

\begin{abstract}

Ferroelectric semiconductors are promising for multifunctional electronics, yet their typically low carrier mobilities remain a major limitation. 
Using first-principles phonon-limited transport calculations for monolayer In$_2$Se$_3$, we show that this limitation depends critically on the microscopic origin of ferroelectricity. 
In displacive $\beta'$-In$_2$Se$_3$, low-frequency ferroelectric modes dominate carrier scattering, with additional contributions from longitudinal-optical (LO) phonons, limiting the room-temperature electron mobility to a few $\cvs$. 
By contrast, in fractional-quantum ferroelectric $\alpha$-In$_2$Se$_3$, ferroelectric-mode scattering is absent because polarization arises from discrete lattice-scale atomic displacements rather than soft-mode condensation. 
Transport is therefore dominated by LO phonons, yielding a room-temperature mobility above 70~$\cvs$. Carrier doping further screens long-range electron--LO-phonon interactions and raises the mobility beyond 300~$\cvs$ at experimentally accessible densities.
These results establish fractional-quantum ferroelectricity can decouple robust polarization from strong intrinsic carrier scattering, offering a route toward high-mobility ferroelectric semiconductors.
\end{abstract}

\newpage
Ferroelectric semiconductors are promising for nonvolatile memory, neuromorphic computing, and photovoltaic applications because they combine switchable polarization with semiconducting transport in a single material~\cite{scott2007applications,setter2006ferroelectric,martin2016thin,chen2016review,khan2020future,kim2023ferroelectric}. Yet their practical use is often limited by a fundamental conflict between ferroelectricity and carrier mobility. In conventional ferroelectrics, the soft phonon modes and anomalously large Born effective charges associated with the ferroelectric instability also generate strong electron-phonon (e-ph) interactions~\cite{cochran1960crystal,zhou2018electron,ma2021large,huang2025coupling,ziffer2022charge}. As a result, charge carriers can be strongly scattered by both low-frequency ferroelectric soft modes and polar longitudinal-optical (LO) phonons, leading to room-temperature mobility far below those of conventional semiconductors~\cite{regmi2016optical,jalandhara2025bifeo3,liu2023optimizing,zhang2021anisotropic,samulionis2001elastic}. Overcoming this intrinsic mobility bottleneck is therefore a central challenge in the development of high-performance ferroelectric semiconductors. \\
\indent
Two-dimensional van der Waals ferroelectrics provide an appealing platform for reexamining this problem. Among them, In$_2$Se$_3$ has emerged as a prototypical system following the prediction of intrinsic ferroelectricity in the $\alpha$ phase~\cite{ding2017prediction} and its subsequent experimental verification~\cite{zhou2017out,xue2018room, poh2018molecular,si2019ferroelectric,han2023phase,zhang2025interlayer}. The $\alpha$-In$_2$Se$_3$ monolayer exhibits a distinctive dipole-locking mechanism, in which out-of-plane polarization is coupled to an in-plane component, enabling robust ferroelectricity down to the monolayer limit~\cite{xiao2018intrinsic,cui2018intercorrelated}. Notably, $\alpha$-In$_2$Se$_3$ field-effect devices have reported room-temperature carrier mobilities exceeding 50~$\cvs$~\cite{han2023phase}, substantially higher than those of most known ferroelectric semiconductors. In$_2$Se$_3$ also hosts a metastable $\beta'$ phase that exhibits conventional displacive ferroelectricity~\cite{zheng2018room,xu2020two} and much lower carrier mobility. 
Despite sharing the same chemical composition and similar layered structures, the two In$_2$Se$_3$ polymorphs exhibit strikingly different transport properties and realize ferroelectricity through distinct microscopic mechanisms, providing an ideal platform for uncovering how the ferroelectric switching mechanism influences carrier transport.\\
\indent
This issue has become especially timely with the recent proposal of fractional-quantum ferroelectricity (FQFE)~\cite{ji2024fractional}. In conventional displacive ferroelectrics, polarization develops through small, continuous ionic displacements associated with soft-mode condensation. By contrast, FQFE arises from discrete atomic displacements on the scale of lattice spacings~\cite{ji2024fractional,yu2025symmetry,pang2025generalized}. The $\alpha$ phase of In$_2$Se$_3$ is the prototypical example: its polarization originates from the hopping of the central Se atom between two symmetry-equivalent positions, rather than from the condensation of a conventional soft ferroelectric mode. This discrete bond-reconfiguration picture suggests that FQFE materials may avoid the strong ferroelectric-mode scattering that limits carrier mobility in displacive ferroelectrics.\\
\indent
In this work, we address this question through \textit{ab initio} calculations of phonon-limited carrier transport in monolayer $\alpha$-In$_2$Se$_3$ and $\beta'$-In$_2$Se$_3$. We find that transport in the two phases is governed by different scattering mechanisms. In displacive $\beta'$-In$_2$Se$_3$, low-frequency ferroelectric modes provide the dominant scattering channel, while LO phonons make an important secondary contribution, together limiting the room-temperature electron mobility to only a few $\cvs$. In contrast, in fractional-quantum ferroelectric $\alpha$-In$_2$Se$_3$, scattering from ferroelectric modes is absent, and carrier transport is governed primarily by polar LO phonons, yielding an intrinsic mobility more than an order of magnitude higher. We further show that carrier doping enhances free-carrier screening of the LO interaction and can raise the room-temperature mobility of $\alpha$-In$_2$Se$_3$ beyond 300~$\cvs$ at experimentally accessible doping levels. These results establish a direct connection between ferroelectric switching mechanism and intrinsic carrier mobility, and identify FQFE as a promising design principle for high-mobility ferroelectric semiconductors.


\begin{figure}
  \includegraphics[width=\textwidth]{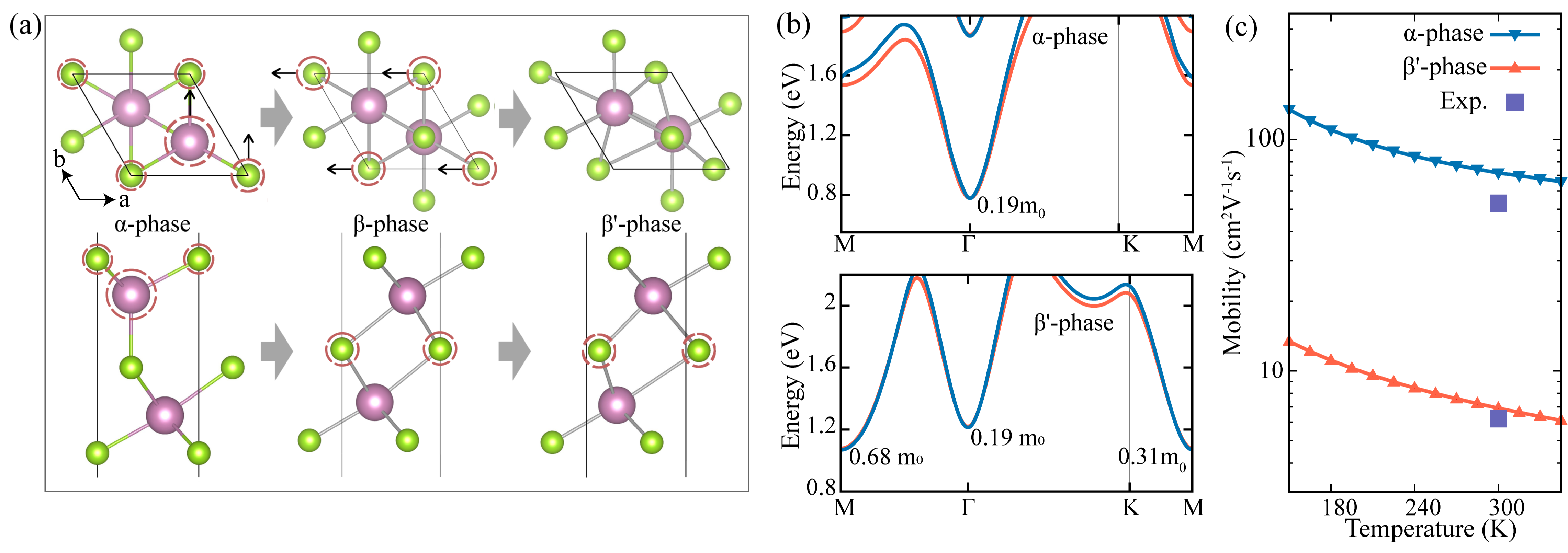}
  \caption{(a) Top and side views of the $\alpha$ phase, the centrosymmetric $\beta$ phase, and the $\beta'$ phase of In$_2$Se$_3$. All three share a quintuple-layer structure (Se$_1$--In$_1$--Se$_2$--In$_2$--Se$_3$) but differ in the coordination environment of the In and Se sublayers. The dashed red circles highlight the atoms displaced during the structural transformations, and the black arrows indicate the corresponding displacement directions. (b) Band structures of the $\alpha$ and $\beta'$ phases computed with the PBE functional (red lines) and $G_0W_0$ corrections (blue lines). The $\alpha$ phase has a $\Gamma$-point conduction band minimum with $m^* = 0.19\,m_0$; the $\beta'$ phase has an M-point minimum with $m^* = 0.68\,m_0$ (M--$\Gamma$) and $0.31\,m_0$ (M--K). (c) Computed phonon-limited electron mobility as a function of temperature, compared with experimental data from Ref.~\citenum{han2023phase}. The computed values along $a$-direction are given for the $\beta^{\prime}$ phase.}
  \label{fig:fig1}
\end{figure}

Figure~\ref{fig:fig1}(a) illustrates the crystal structures of three representative In$_2$Se$_3$ polymorphs. All three share a common quintuple-layer architecture (Se$_1$--In$_1$--Se$_2$--In$_2$--Se$_3$) 
but differ in the relative arrangement of these atomic planes. The $\alpha$ phase (space group $P3m1$) is characterized by an asymmetric coordination environment in which one indium sublayer is tetrahedrally coordinated while the other is octahedrally coordinated, generating a net electric dipole. The centrosymmetric $\beta$ phase ($P\bar{3}m1$) is obtained by translating the topmost In$_1$ and Se$_1$ atoms by $(1/3)\bm{a} + (2/3)\bm{b}$, which aligns In$_1$ vertically above the bottom Se$_3$ atom and restores inversion symmetry. The $\beta'$ phase is derived from $\beta$ by a small in-plane displacement of the central Se$_2$ atom from its high-symmetry position.\\
\indent
The $\alpha$ and $\beta'$ phases realize ferroelectricity and ferroelectric switching through fundamentally distinct mechanisms. In the $\beta'$ phase, ferroelectricity arises from small, continuous displacements of the central Se$_2$ atom that break inversion symmetry, reflecting a classic soft-mode instability in which a low-frequency transverse optical phonon condenses to produce spontaneous polarization~\cite{tang2022strain}. 
In the $\alpha$ phase, the central Se$_2$ atom occupies one of three high-symmetry sites within each layer. When this atom undergoes a lateral displacement to an adjacent high-symmetry site related by inversion symmetry, both the in-plane and out-of-plane polarization components reverse, which is a consequence of the dipole-locking mechanism intrinsic to this structure. Because the initial and final states are related by a fractional lattice translation, the polarization difference between them is a fractional multiple of the polarization quantum $\bm{Q} = e\bm{R}/\Omega$, making $\alpha$-In$_2$Se$_3$ a prototype fractional-quantum ferroelectric~\cite{ji2024fractional}. Crucially, the transition is a discrete structural rearrangement over a finite energy barrier ($\sim$68~meV per unit cell)~\cite{ji2024fractional}, not a continuous condensation of a zone-center soft phonon.\\
\indent
The electronic band structures of both phases, computed at the PBE and $G_0W_0$ levels, are shown in Figure~\ref{fig:fig1}(b). The $\alpha$ phase has a conduction band minimum (CBM) at $\Gamma$ with an isotropic effective mass of $m^* = 0.19\,m_0$. The $\beta'$ phase has its CBM at the M point with anisotropic effective masses of $0.68\,m_0$ along M--$\Gamma$ and $0.31\,m_0$ along M--K, both heavier than in the $\alpha$ phase. Notably, the $\beta'$ conduction band also possesses a secondary minimum at $\Gamma$, about 140~meV above the CBM, with the same effective mass  as the $\alpha$ phase.\\
\indent
We perform first-principles calculations of phonon-limited electron mobility for both the $\alpha$ and $\beta'$ phases in the intrinsic or lightly doped regime where free-carrier screening effects are neglected (see Methods). Figure~\ref{fig:fig1}(c) shows the computed electron mobility as a function of temperature. At 300~K, we obtain $\mu_\alpha$~=~72~cm$^2$/Vs for the $\alpha$ phase, while the $x$-- and $y$--direction mobilities of the $\beta^{\prime}$ phase are 6.3~cm$^2$/Vs and 6.9~cm$^2$/Vs, respectively, with the $x$ direction defined along the $a$ axis and the $y$ direction perpendicular to the $a$ axis [see Figure~1(a)]. The $x$-direction mobility ratio between the two phases is $\mu_\alpha/\mu_{\beta^{\prime}}\approx11$.
Both values agree well with experimental results, validating our first-principles calculations and confirming that the mobility contrast is intrinsic rather than limited by extrinsic factors such as defects.\\
\indent
The heavier effective masses at the CBM of the $\beta'$ phase account for part of this order-of-magnitude mobility contrast, but not all of it. 
Within a simple Drude picture ($\mu = e\tau/m^*$), the mass difference alone would yield a mobility ratio of two to three; matching the observed factor of eleven requires the electron relaxation time in the $\alpha$ phase to substantially exceed that in the $\beta'$ phase.
To quantify the respective contributions, we perform a quantitative factorization analysis of the electron mobility within the relaxation-time approximation~\cite{pant2026gwpt} (see Table~S1 and the accompanying discussion in Item 9 of the Supporting Information). The difference in band structure contributes a factor of 2.67, while the difference in relaxation time contributes a factor of 4.11.
This points to a fundamental difference in e-ph scattering mechanisms between the two phases.

\begin{figure}
  \includegraphics[width=\textwidth]{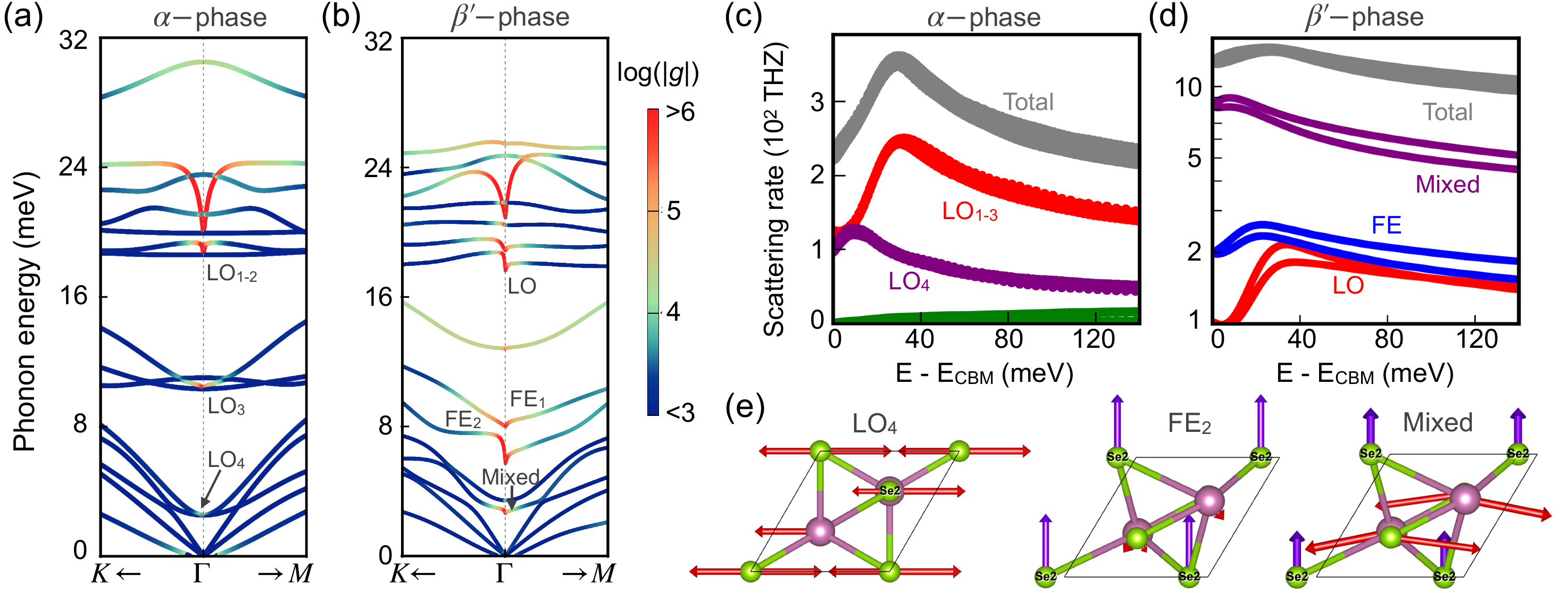}
  \caption{Mode-resolved e-ph coupling and scattering rates in the $\alpha$ and $\beta'$ phases. (a,b) Phonon dispersions overlaid with a logarithmic color map of the e-ph coupling strength $|g|$, evaluated for initial electron states at $\Gamma$ ($\alpha$ phase) and M ($\beta'$ phase). In the $\alpha$ phase, coupling is concentrated on the LO branches; in the $\beta'$ phase, additional strong coupling appears on the low-frequency ferroelectric modes (FE$_1$, FE$_2$) and a hybrid mode (Mixed) that combines ferroelectric and LO character. (c,d) Energy-resolved e-ph scattering rates at 300~K for the $\alpha$ and $\beta'$ phases, decomposed by phonon branch. In (c), scattering is dominated by LO phonons. In (d), ``LO'' denotes the sum over three LO branches, ``FE'' the sum over the FE$_1$ and FE$_2$ ferroelectric modes, and ``Mixed'' the hybrid mode identified in (b). For clarity, only scattering rates for electronic states along K--$\Gamma$ and $\Gamma$--M are shown. (e) Atomic displacement patterns of the LO$_4$ mode in the $\alpha$ phase and the FE$_2$ and Mixed modes in the $\beta'$ phase, illustrating the distinct role of the central Se$_2$ atom in each case.}
\end{figure}

To elucidate the microscopic origin of the mobility contrast identified above, we perform a mode-resolved analysis of e-ph coupling in both phases. Figures~2(a) and 2(b) display the phonon dispersions of the $\alpha$ and $\beta'$ phases overlaid with a logarithmic color map of the mode-resolved coupling strength $|g_{\nu}(\bm{q})|$ (see Methods). In the $\alpha$ phase [Figure~2(a)], conduction electrons at the $\Gamma$-point CBM couple predominantly to four LO branches (LO$_1$--LO$_4$) via long-range Fr\"ohlich interactions, with the coupling strength concentrated near $\Gamma$ and diminishing at larger momenta. This coupling pattern is typical of conventional polar semiconductors such as GaAs or SrSnO$_3$, with no indication of strong low-frequency scattering channels~\cite{zhou2016ab,liu2017first,truttmann2021combined}.

The $\beta'$ phase presents a qualitatively different picture. In addition to three LO branches, its low-frequency spectrum contains modes that originate from the structural instability of the parent $\beta$ phase. The centrosymmetric $\beta$ phase possesses a doubly degenerate unstable mode at $\Gamma$, corresponding to in-plane displacement of the central Se$_2$ atom (Figure~S1). Upon condensation into the $\beta'$ phase, Se$_2$ shifts along a specific in-plane direction, breaking the in-plane isotropy and lifting this degeneracy into two distinct real-frequency modes, FE$_1$ and FE$_2$, whose eigenvectors remain dominated by Se$_2$ in-plane motion (Figure~S1). More broadly, the off-centering of Se$_2$ modifies its participation in all phonon eigenvectors, not only the ferroelectric modes. In particular,  a third low-frequency mode (labeled ``Mixed'') combines a large in-plane Se$_2$ displacement resembling the ferroelectric distortion with polar LO-like motion of the remaining atoms [Figure~2(e)], producing strong e-ph coupling.

The energy-resolved scattering rates at 300~K, shown in Figure~2(c) and 2(d), reveal how these coupling patterns translate into carrier scattering. Starting with the $\alpha$ phase [Figure~2(c)], the total scattering rate is almost entirely accounted for by the four LO branches. Among these, LO$_4$ makes a disproportionately large contribution despite having the weakest e-ph coupling strength $|g_\nu|$: its lower phonon frequency yields a larger thermal occupation at room temperature and a lower emission threshold that allows scattering to begin at smaller electron energies, amplifying the scattering rate (Figure~S2 in Supporting Information).
Crucially, beyond these LO branches, no low-frequency mode contributes appreciably to scattering.
This is a direct consequence of the fractional-quantum ferroelectric mechanism, in which polarization originates from discrete lattice-scale atomic displacements rather than soft-mode condensation.

The scattering landscape in the $\beta'$ phase is fundamentally different [Figure~2(d)]. The Mixed mode alone produces a scattering rate approximately four times that of all three LO branches combined, while the ferroelectric modes FE$_1$ and FE$_2$ each scatter carriers more efficiently than any individual LO mode. 
%
To quantify the contributions of different phonon-scattering channels to the calculated mobility, we recalculated the electron mobility of $\beta'$-In$_2$Se$_3$ after selectively excluding each channel while keeping the others unchanged (Table~S2 in the Supporting Information). We find that the Mixed mode has the largest individual impact, with its removal nearly doubling the mobility relative to the all-channel result. Excluding the FE$_1$, FE$_2$, and Mixed modes together increases the mobility to roughly three times the all-channel value, demonstrating that their combined scattering substantially limits mobility in $\beta'$-In$_2$Se$_3$.
This hierarchy, in which soft ferroelectric modes and their hybrids dominate over conventional polar-optical scattering, is the hallmark of displacive ferroelectrics and constitutes the primary microscopic origin of mobility suppression in the $\beta'$ phase.

The enhanced scattering in the $\beta^{\prime}$ phase can be mainly traced to the stronger e-ph coupling of the Mixed mode, as reflected by its larger mode-effective charge.
The mode-effective charge combines the Born effective-charge tensors with the phonon eigen-displacements and directly governs the long-range e-ph coupling (Item~3 in Supporting Information).
Near $\Gamma$, the mode-effective charge of the $\beta^{\prime}$-phase Mixed mode is more than five times larger than that of the $\alpha$-phase LO$_4$ mode.
This enhancement originates from both the nearly twofold larger Born effective charge of the central Se$_2$ atom in $\beta^{\prime}$-In$_2$Se$_3$, associated with its displacive ferroelectric instability, and the distinct phonon displacement patterns.
An algebraic decomposition of the long-range e-ph coupling using the Born effective charges and eigenvectors of the full crystal shows that the Se$_2$ contribution constructively enhances the long-range coupling of the Mixed mode, whereas it partially cancels the coupling of the LO$_4$ mode in $\alpha$-In$_2$Se$_3$ (Figure~S3 in Supporting Information).
The combined effects of the enhanced Born effective charge and favorable displacement pattern therefore produce a substantially larger mode-effective charge and stronger e-ph scattering in the $\beta^{\prime}$ phase.

These results establish a direct link between the type of ferroelectric order and intrinsic carrier transport. 
In displacive ferroelectric $\beta'$-In$_2$Se$_3$, the soft polar phonon that underpins ferroelectricity simultaneously acts as the dominant carrier scattering channel, creating an intrinsic conflict between robust polarization and high mobility. In fractional-quantum ferroelectric $\alpha$-In$_2$Se$_3$, no such conflict exists: the polarization arises from discrete atomic rearrangements that do not produce a soft mode, leaving only conventional LO-phonon scattering to limit transport. This decoupling of ferroelectric order from soft-mode scattering is a primary factor why the $\alpha$ phase achieves an order-of-magnitude higher mobility than the $\beta'$ phase.

\begin{figure}
  \includegraphics[width=0.6\columnwidth]{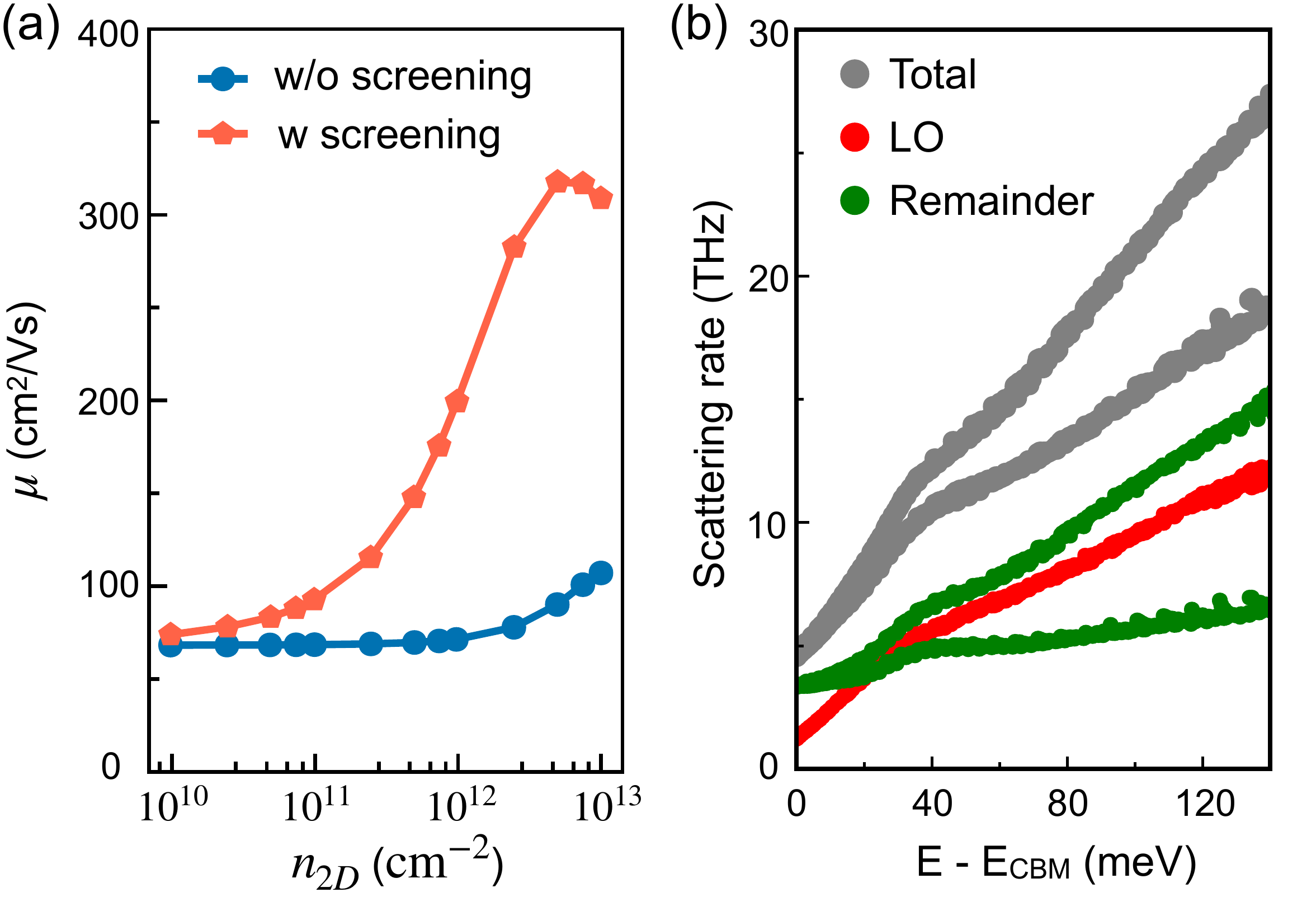}
  \caption{Effect of free-carrier screening on electron mobility in the $\alpha$ phase. (a) Room-temperature electron mobility as a function of carrier concentration, computed with (red) and without (blue) free-carrier screening of e-ph interactions. Screening enhances the mobility from $\sim$72 to $\sim$318~$\cvs$ at an optimal doping of $\sim$$5 \times 10^{12}$~cm$^{-2}$. (b) Energy-resolved scattering rates at 300~K and $n_\mathrm{2D} = 5 \times 10^{12}$~cm$^{-2}$, showing that LO-phonon scattering (``LO'', summed over all LO branches) is strongly suppressed by screening compared with the undoped case [Figure~2(c)].}
\end{figure}

We further show that the phonon-limited mobility of $\alpha$-In$_2$Se$_3$ can be substantially enhanced by free-carrier screening at experimentally accessible doping levels.
Figure~3(a) shows the room-temperature electron mobility of $\alpha$-In$_2$Se$_3$ as a function of carrier concentration, computed with and without free-carrier screening of e-ph interactions, as described in Methods.
Without free-carrier screening, the electron mobility increases only weakly with carrier concentration, reaching $\sim$107~cm$^2$\,V$^{-1}$\,s$^{-1}$ at $n_c=1.0\times10^{13}$~cm$^{-2}$.
This weak increase arises from the competing changes in carrier occupation and scattering phase space with doping.
With free-carrier screening, the mobility increases slowly at low carrier concentration, where screening remains weak; it then rises rapidly at intermediate concentrations, reaching a peak of $\sim$318~cm$^2$\,V$^{-1}$\,s$^{-1}$ at $n_c \approx 5\times10^{12}$~cm$^{-2}$,  a more than fourfold enhancement. At higher concentrations the mobility decreases slightly, as the screening benefit saturates while the enlarged scattering phase space begins to dominate.
We note that the optimal doping concentration of $n_c \approx 5 \times 10^{12}$~cm$^{-2}$ is well within the range achieved in $\alpha$-In$_2$Se$_3$ field-effect devices, where carrier densities exceeding $10^{13}$~cm$^{-2}$ have been demonstrated through electrostatic gating~\cite{si2019ferroelectric}.

This enhancement originates from the free-carrier screening of the long-range electron--LO phonon interaction, which suppresses the dominant scattering channel of $\alpha$-In$_2$Se$_3$.
This screening effect is apparent in the mode-resolved e-ph coupling strengths at representative carrier concentrations (see Figure~S4).
With increasing concentration, the coupling between electron and long-wavelength LO phonon is progressively suppressed.
Meanwhile, the coupling between electron and TO phonon barely changes at different doping, because it arises from short-range interactions and is insensitive to screening.
These changes in the e-ph coupling are directly reflected in the scattering rates. We split the total scattering rate at $n_c = 5\times10^{12}$~cm$^{-2}$ into a long-range LO-phonon contribution and the remaining contributions, as shown in Figure~3(b). The LO-phonon scattering rate is strongly suppressed compared with the undoped case [Figure~2(c)], and becomes comparable with the remainder. 

\indent
We also performed phonon-limited transport calculations including free-carrier screening for the $\beta'$ phase (Figure~S5).
The mobility enhancement induced by doping is much weaker in the $\beta'$ phase than in the $\alpha$ phase: even at $n_c \approx 10^{13}$~cm$^{-2}$, the increase is only about 69\% relative to the intrinsic mobility.
The smaller mobility enhancement in $\beta^{\prime}$-In$_2$Se$_3$ arises primarily from substantial residual short-range and finite-$q$ coupling to low-frequency FE-derived modes, despite comparable free-carrier dielectric screening in the two phases. These modes remain effective scattering channels after screening, whereas the dominant long-wavelength Fr\"ohlich coupling in $\alpha$-In$_2$Se$_3$ is strongly suppressed (see Figure~S7 and the accompanying discussion in Item 7 of the Supporting Information).

We assessed the effect of carrier screening on the stability of the ferroelectric structures by following the evolution of the relaxed structures under electron doping (Fig. S8 in the Supporting Information). The two phases exhibit markedly different responses: the Se$_2$ off-centering in $\beta'$-In$_2$Se$_3$ is progressively suppressed with increasing carrier density, with the structure becoming nearly paraelectric at $5\times10^{14}$~cm$^{-2}$, whereas the $\alpha$-In$_2$Se$_3$ remains essentially unchanged over the studied range.
This contrast further underscores the advantage of the FQFE.
Our results suggest that moderately doped FQFE materials offer a practical route toward ferroelectric semiconductors whose room-temperature mobilities can approach those of established nonpolar two-dimensional channel materials.

We emphasize that these mobilities calculated in our work should be regarded as phonon-limited upper bounds rather than expected device mobilities, since other scattering mechanisms, including device-dependent scattering from dielectric charged impurities, interface disorder, and remote-interface phonons, are not considered in our calculations.
In addition, polarization reversal may involve correlated atomic motion, domain nucleation, and domain-wall propagation, generating slow or quasi-static disorder that can also scatter carriers. A quantitative assessment of these switching-related scattering processes is beyond the scope of the present work and remains for future investigation.


In summary, we have performed first-principles phonon-limited transport calculations for two monolayer In$_2$Se$_3$ polymorphs that realize ferroelectricity through fundamentally different mechanisms. Our results reveal a direct connection between the microscopic origin of ferroelectric order and intrinsic carrier mobility. In displacive ferroelectric $\beta'$-In$_2$Se$_3$, low-frequency ferroelectric phonon modes and their hybrids with LO phonons dominate electron scattering, limiting the room-temperature mobility to $\sim$6~$\cvs$. In fractional-quantum ferroelectric $\alpha$-In$_2$Se$_3$, these soft-mode scattering channels are absent because polarization arises from discrete lattice-scale atomic rearrangements rather than soft-mode condensation. Carrier transport is instead governed by conventional LO-phonon Fr\"ohlich scattering, yielding an intrinsic mobility of $\sim$72~$\cvs$, more than an order of magnitude higher than that of the $\beta'$ phase. Free-carrier screening at experimentally accessible doping levels further suppresses the LO-phonon interaction and raises the mobility by more than fourfold. These findings establish that fractional-quantum ferroelectricity can decouple robust polarization from strong intrinsic carrier scattering, resolving a long-standing conflict that has limited the performance of ferroelectric semiconductors.

Our results point to several promising directions. 
The mechanism identified here suggests that FQFE materials avoiding soft-mode condensation may be promising candidates for high-mobility ferroelectric semiconductors. 
The recent discovery of over 200 FQFE candidates therefore provides a broad materials space for testing this possibility through future transport calculations.~\cite{yu2025symmetry}. 
Beyond materials discovery, the effectiveness of free-carrier screening demonstrated for $\alpha$-In$_2$Se$_3$ motivates the exploration of device architectures that combine electrostatic gating with FQFE channels, as well as heterostructure approaches in which remote metallic layers provide additional screening of the Fr\"ohlich interaction~\cite{sohier2021remote,sohier2017density,wu2023electrostatic,radisavljevic2011single}. More broadly, the same principle may extend to sliding ferroelectrics, a rapidly growing family of two-dimensional van der Waals materials in which polarization is switched by lateral interlayer translation~\cite{li2017binary,wu2021sliding}. Like fractional-quantum ferroelectrics, sliding ferroelectrics such as bilayer hBN~\cite{yasuda2021stacking,vizner2021interfacial} and rhombohedral-stacked transition metal dichalcogenides~\cite{wang2022interfacial} switch through discrete structural rearrangements with no soft polar phonon involved, suggesting that they too may avoid the intrinsic mobility penalty of displacive ferroelectricity.
Taken together, these considerations suggest that the absence of soft-mode condensation may serve as a broadly applicable design criterion for identifying ferroelectric semiconductors with high intrinsic carrier mobility.

\section{Methods}

\textbf{DFT, DFPT and GW calculations.}
First-principles calculations were performed using the \textsc{Quantum ESPRESSO} code, employing the PBE exchange-correlation functional and the Optimized Norm-Conserving Vanderbilt Pseudopotential from SG15-1.2 \cite{giannozzi2017advanced,giannozzi2009quantum,hamann2013optimized}. Calculations utilized a $\bm{k}$-point grid of $16\times16\times1$, a plane wave cutoff of 95 Ry, a vacuum layer thickness of 20~$\text{\AA}$, and a cutoff for Coulomb interactions in the out-of-plane direction following the method of Sohier et al.~\cite{sohier2017density}. One-shot $G_0W_0$ calculations were performed using the \textsc{Yambo} code \cite{marini2009yambo,sangalli2019many}. A cutoff energy of 12~Ry was employed for the dielectric screening, together with the Bruneval--Gonze terminator \cite{bruneval2008accurate}. The response functions were evaluated using 280 bands and a $28 \times 28 \times 1$ $\bm{k}$-point grid \cite{aryasetiawan1998gw}; we have verified that increasing the number of bands to 360 and the dielectric cutoff energy to 16~Ry has a negligible effect.
Phonon dispersions and perturbation potentials were obtained using density functional perturbation theory (DFPT) on a coarse $8 \times 8 \times 1$ $\bm{q}$-point grid~\cite{baroni2001phonons}. The carrier mobility as a function of temperature and doping concentration was obtained by iteratively solving the Boltzmann transport equation on dense $500 \times 500 \times 1$ $\bm{k}$- and $\bm{q}$-point grids using the \textsc{Perturbo} package~\cite{zhou2021perturbo}.
The PBE electronic energies are used in the BTE transport calculations.  The $G_0W_0$ correction only has a small correction to the conduction band effective masses [see Figure~1(b)]; we expect it to have a small impact on the transport results.\\
\newline

\textbf{E-ph coupling matrix elements.}
The e--ph matrix element \cite{bernardi2016first}
\begin{equation}
	g_{mn\nu}(\bm{k}, \bm{q})=
	\sqrt{\frac{\hbar}{2\omega_{\nu\bm{q}}}}
	\sum_{\kappa\alpha}
	\frac{e^{\kappa\alpha}_{\nu\bm{q}}}{\sqrt{M_{\kappa}}}
	\left\langle
	\psi_{m\bm{k}+\bm{q}}
	\left|
	\partial_{\bm{q},\kappa\alpha} V
	\right|
	\psi_{n\bm{k}}
	\right\rangle
\end{equation}
describes the scattering of an electron from Bloch state $|n\bm{k}\rangle$ into $|m\bm{k}+\bm{q}\rangle$ via emission or absorption of a phonon with frequency $\omega_{\nu\bm{q}}$. Here $M_{\kappa}$ is the mass of atom $\kappa$, and $e^{\kappa\alpha}_{\nu\bm{q}}$ denotes the phonon eigenvector for atom $\kappa$ along the Cartesian direction $\alpha$.

Figure~2 shows the phonon dispersion overlaid with a color map of the mode-resolved e-ph coupling strength
\begin{equation}
	|g_{\nu}(\bm{q})| = \left[\sum_{mn} |g_{mn\nu}(\bm{k}_0, \bm{q})|^2 / N_b\right]^{1/2},
\end{equation}
where $N_b$ is the number of selected bands and $\bm{k}_0$ is taken at the CBM of each phase ($\Gamma$ for the $\alpha$ phase and $M$ for the $\beta'$ phase).

Following previous works \cite{verdi2015frohlich,jhalani2020piezoelectric,park2020long,Brunin2020}, the total e--ph coupling strength $g$ can be decomposed into a short-range component $g^{\mathrm{S}}$ and a long-range component $g^{\mathrm{L}}$:
\begin{equation}
	g = g^{\mathrm{S}} + g^{\mathrm{L}} = g^{\mathrm{S}} + g^{\mathrm{dip}} + g^{\mathrm{quad}} ,
\end{equation}
where $g^{\mathrm{dip}}$ and $g^{\mathrm{quad}}$ denote the dipole and quadrupole contributions, respectively. The long-range matrix elements are evaluated within a two-dimensional framework that incorporates both dipole and quadrupole contributions, using a range-separation technique to partition real-space and reciprocal-space treatments \cite{royo2021exact, ponce2023accurate}. The dynamical quadrupoles are computed using the \textsc{ABINIT} code and converted into their effective 2D form following the procedure described in Ref.~\citenum{royo2021exact}.\\
\newline

\textbf{Free-carrier screening of e-ph coupling.}
In doped semiconductors, free carriers partially screen the macroscopic electric fields generated by the dipole and quadrupole responses in Eq.~(3), thereby reducing the strength of long-range e--ph coupling. In our treatment, we assume that only the long-range component $g^{\mathrm{L}}$ is affected by free-carrier screening, while the short-range component $g^{\mathrm{S}}$ remains unchanged. This is physically motivated by the fact that $g^{\mathrm{S}}$ arises from local deformation-potential interactions that do not produce macroscopic electric fields and are therefore insensitive to electrostatic screening \cite{macheda2022electron,macheda2023electron}. Accordingly,
\[
g = g^{\mathrm{S}}+g^{\mathrm{L}}
\quad \rightarrow \quad
g = g^{\mathrm{S}}+g^{\mathrm{L},\mathrm{dop}}.
\]

The free-carrier screening enters through the modification of the independent-particle polarizability, $\chi_{0}\rightarrow\chi_{0}+\delta\chi_{0}$, which modifies the in-plane 2D dielectric function within the random-phase approximation (RPA) as
\begin{equation}
	\epsilon^{\|}(\bm{q})
	\rightarrow
	\epsilon^{\|}(\bm{q})
	-
	v\,\delta\chi_{0},
\end{equation}
where $v=2\pi/|\bm{q}|$ is the 2D Coulomb interaction. The correction $\delta\chi_{0}$ is given by \cite{macheda2023electron}
\begin{equation}
	\begin{aligned}
		\delta\chi_{0}(\bm{q},n,T)
		=&
		\frac{2e^{2}}{A}
		\sum_{mm'\bm{k}}
		\frac{\delta f_{m\bm{k}}-\delta f_{m'\bm{k}+\bm{q}}}
		{\varepsilon_{m\bm{k}}-\varepsilon_{m'\bm{k}+\bm{q}}} \\
		&\times
		\left|
		\left\langle
		u_{m\bm{k}}
		\mid
		u_{m'\bm{k}+\bm{q}}
		\right\rangle
		\right|^{2},
	\end{aligned}
\end{equation}
where $A$ is the unit-cell area,
$\delta f_{m\bm{k}} = f_{m\bm{k}}^{\mathrm{dop}} - f_{m\bm{k}}^{\mathrm{undop}}$, and $f$ is the Fermi--Dirac distribution function. $\varepsilon_{m\bm{k}}$ denotes the band energy, and $u$ represents the periodic part of the Bloch wave function.
Note that the free-carrier screening is treated within the static approximation; therefore dynamical screening effects, such as plasmon-phonon hybridization, are not included in our calculations.
\newline

\begin{acknowledgement}
The authors thank the National Natural Science Foundation of China (Grant No. 12574250 and No. 12104039),
the Beijing Natural Science Foundation (Grant No. Z210006),
and the National Key R\&D Program of China (Grant No. 2022YFA1403400) for financial support.
\end{acknowledgement}

\begin{suppinfo}
Additional information on the phonon dispersions, atomic displacement patterns, e--ph coupling, and scattering rates is provided in the Supporting Information.
\end{suppinfo}

\section*{Notes}
The authors declare no competing financial interest.


\bibliography{ref}

\end{document}